\documentclass[%
 reprint,
 amsmath,amssymb,
pra,
]{revtex4-2}

\usepackage{graphicx}
\usepackage{dcolumn}
\usepackage{bm}

\begin{document}


\title{Real-time observation of single atoms trapped and interfaced to a nanofiber cavity}

\author{Kali P. Nayak$^\dagger$$^\star$}

 \author{Jie Wang$^\dagger$$^\ddagger$}
\author{Jameesh Keloth}
\affiliation{Center for Photonic Innovations and Institute for Laser Science, University of Electro-Communications, Chofu-shi, Tokyo 182-8585}


\begin{abstract}
We demonstrate an optical tweezer based single atom trapping on an optical nanofiber cavity. We show that the fluorescence of single atoms trapped on the nanofiber cavity can be readily observed in real-time through the fiber guided modes. The photon correlation measurements further clarify the atom number and the dynamics of the trap. The trap lifetime is measured to be 52$\pm$5 ms. From the photon statistics measured for different cavity decay rates, the effective coupling rate of the atom-cavity interface is estimated to be 34$\pm$2 MHz. This yields a cooperativity of 5.4$\pm$0.6 and a cavity enhanced channeling efficiency of 85$\pm$2\% for a cavity mode having a linewidth of 164 MHz. These results may open new possibilities for building trapped single atom based quantum interfaces on an all-fiber platform.  

\end{abstract}

\maketitle



Realizing an efficient quantum interface will enable deterministic control and readout of the quantum states for quantum information processing \cite{kimble}. In this context, atomic qubits offer unique capabilities for long coherence time and optical interfacing \cite{kimble, remperev}. A recent trend towards optical interfacing is to combine the atomic qubits with nanophotonic platforms and develop hybrid quantum systems, where efficient quantum state-transfer between atoms and photons can be realized.

Adiabatically tapered single mode optical fiber with subwavelength diameter waist, referred to as optical nanofiber (ONF), provides a unique fiber-in-line platform for quantum photonics applications \cite{famsan1, CPIreview}. The transverse confinement of photonic modes in the ONF has enabled new possibilities for manipulating atom-photon interactions \cite{famsan1, kpn1, arno0, CPIreview}. 
Furthermore, it has been demonstrated that thousands of atoms can be trapped in the vicinity of the ONF by using guided light in a two-color dipole trap scheme \cite{tfam1, tfam2, arno1}. This inherently leads to a fiber-in-line optically dense platform for quantum non-linear optics with an ensemble of atoms \cite{arno2,julien}. 

On the other hand, for achieving non-linearity at single atom level, the coupling between the atom and the ONF guided modes can be substantially improved by introducing an in-line fiber cavity. Due to the strong transverse confinement of ONF guided modes, one can achieve strong-coupling regime of cavity quantum electrodynamics (QED) and high single atom cooperativity even for a moderate finesse ONF cavity \cite{famsan2}. 
A detailed review of various types of ONF cavities, can be found in Ref. \cite{CPIreview}. 
Recently, strong-coupling between a trapped single atom and an ONF cavity has been demonstrated \cite{aoki}. However, the widely adopted scheme of two-color guided mode trapping of atoms on the ONF \cite{tfam1, tfam2, arno1, arno2, julien, aoki}, is the only scheme so far and it lacks the control over individual atoms. Preparation of quantum emitters on the ONF cavity, in a bottom-up approach using atom-by-atom control may open new possibilities to engineer complex quantum systems. Therefore, development of new trapping schemes for atom-ONF interface is essential.  

Here, we demonstrate an optical tweezer based side-illumination trapping scheme to trap and interface individual single atoms to an ONF cavity. We show that the fluorescence of single atoms trapped on the ONF cavity can be readily observed in real-time through the fiber guided modes. The atom number and the dynamics of the trap are further clarified from the photon correlation measurements. The trap lifetime is measured to be 52$\pm$5 ms. From the photon statistics measured for different cavity decay rates, the effective coupling rate of the atom-cavity interface is estimated to be 34$\pm$2 MHz. This yields a cooperativity of 5.4$\pm$0.6 and a cavity enhanced channeling efficiency of 85$\pm$2\% for a cavity mode having a linewidth of 164 MHz. These results may open new possibilities for building trapped single atom based quantum interfaces on an all-fiber platform.

\begin{figure}[tbp]
\includegraphics[width=\linewidth]{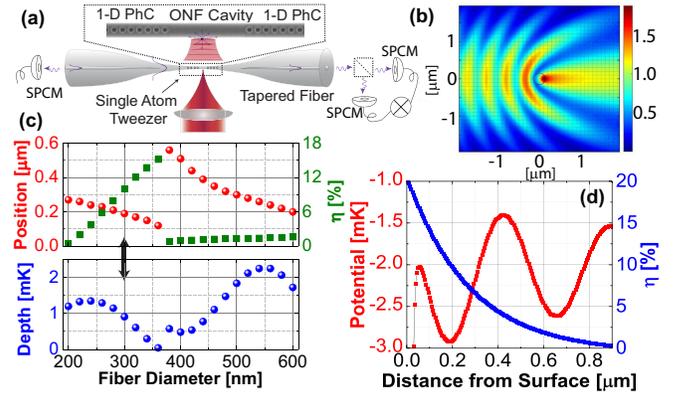}
\caption{\label{fig1} (Color Online) Optical tweezer based side-illumination ONF trap. (a) Schematic diagram of the experiment. (b) Intensity distribution of the tweezer beam around the ONF estimated using FDTD simulation. 
(c) The upper panel shows the estimated trap position (red dots) from the fiber surface and the corresponding channeling efficiency ($\eta$) (green squares) for different fiber diameters. The lower panel shows the corresponding trap depth estimated assuming a beam waist of 1 $\mu$m and a power of 14 mW. (d) The estimated trapping potential (red squares) and the corresponding $\eta$ (blue squares) plotted as a function of radial separation from the fiber surface, for a fiber diameter of 300 nm (indicated by the black arrow in (c)).}
\end{figure}

A schematic diagram of the experiment is shown in Fig. 1(a). An ONF cavity is formed by fabricating two photonic crystal (PhC) structures on the ONF using femtosecond laser ablation \cite{phcn1, phcn2, LPhCNF}. Single atoms are trapped on the ONF segment between the two PhCs. The dipole trap beam is tightly focused on the nanofiber and forms an optical tweezer for single atoms. When the tweezer beam hits the ONF, it forms standing-wave like nanotraps close to ONF due to the interference of the incident and the reflected lights from the fiber surface. The fluorescence of trapped single atom coupled to the ONF cavity mode are detected at the either ends of the fiber using single photon counting modules (SPCMs).

Figure 1(b) shows the typical intensity distribution of the tweezer beam around the ONF estimated using finite difference time domain (FDTD) simulation. The black circle at the center indicates the ONF position. The ONF diameter used for the simulation is 300 nm. The tweezer beam having a wavelength of 938 nm (red-detuned magic wavelength for Cs-atom \cite{tfam2}) is focused to 1 $\mu$m beam waist and is irradiated perpendicular to the ONF from the left side. The polarization is chosen to be parallel to the fiber axis. It may be seen that standing-wave like intensity pattern is formed in the irradiation-side, resembling an 1-D optical lattice. The red-detuned trapping light forms trapping minimum at the maximum intensity location. Therefore the first lattice site nearest to the fiber surface enables tight localization of the trap along the radial direction of the ONF. The tightly focused tweezer beam enables localization of the trap along the axial and azimuthal direction of the ONF.  

For a given trapping laser wavelength, the position of the first lattice site depends on the fiber diameter. 
The channeling efficiency ($\eta$) of spontaneous emission of atom into ONF guided modes (a measure of the interaction strength) also depends on the ONF diameter and the radial position of atom from the ONF \cite{famsan1}. Therefore the selection of ONF diameter, is a crucial design principle for the efficient interfacing of the trapped atom with the ONF guided modes. 

The upper panel in Fig. 1(c), shows the estimated trap position (red dots) from the fiber surface and the corresponding $\eta$ (green squares) for different fiber diameters. The $\eta$ is estimated for the excited state ($6P_{3/2}|F=5,m_F=5\rangle$) of the Cs-atom D2-transition \cite{famsan1}. The lower panel shows the corresponding trap depth estimated assuming a beam waist of 1 $\mu$m and a power of 14 mW. The contribution from the van der Waals (vdW) potential is included for the estimation of the trap depth \cite{tfam1}. It can be seen that for the diameter range of 300-350 nm, the trap position can be 100-200 nm from the fiber surface and the $\eta$ at the trap position can be as high as 10-15\%. For much higher or lower diameters the $\eta$ can be much lower. Moreover for a diameter close to 350 nm, although the $\eta$ is highest but the trap depth is smallest. It is due to the fact that the trap position is close to 100-150 nm and the vdW potential strongly affects the trap depth. Therefore for the experiments we have chosen a fiber diameter around 300 nm. 

Figure 1(d) shows the estimated trapping potential and the corresponding channeling efficiencies for a 300 nm diameter fiber, plotted as a function of radial distance from the fiber surface. It may be seen that the closest trapping minimum is created at $\simeq 190$ nm from the fiber surface with a trap depth of $\simeq 0.9$ mK and a $\eta$ of 10\% can be realized at the trap position. The trap frequencies along the radial and axial direction of the ONF are estimated to be 380 and 80 kHz, respectively. It should be noted that the $\eta$ at the second lattice site is only 1.4\%. 


\begin{figure}[tbp]
\includegraphics[width=\linewidth]{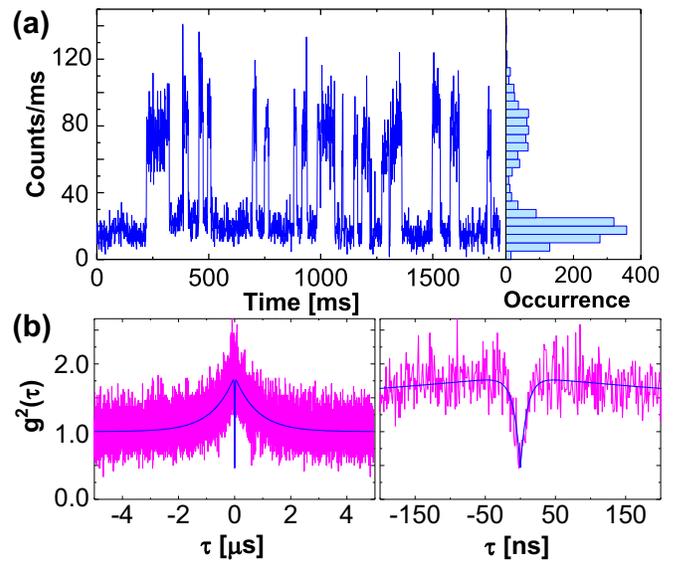}
\caption{\label{fig2} (Color Online) (a) Real-time observation of trapped single atoms. The blue trace in the left panel shows the typical photon counts measured through the ONF cavity. The right panel shows the histogram of the photon counts. (b) Photon correlation of the ONF trap signal. The magenta trace shows the normalized correlation (g$^2$($\tau$)) of the photon counts measured through the ONF cavity. The blue curve shows the theoretical fit. The right panel shows the enlarged view of the plot shown in the left panel.}
\end{figure}

In the experiment, we have used an ONF cavity that has a central ONF segment with a diameter of 300$\pm$10 nm. From the measured free spectral range ($\Delta \nu_{FSR} =23 \pm 2$ GHz), the optical length ($L$) of the cavity is estimated to be 6.5$\pm$0.5 mm. The ONF cavity modes have linewidths in the range 150-1200 MHz and can be tuned to the Cs-atom resonance by stretching the tapered fiber using a piezo actuator (Attocube, ANPx51) attached to the fiber holder. The dipole trap beam is tightly focused to 1 $\mu$m beam waist using a high numerical aperture ($NA=0.5$) lens introduced into the vacuum chamber, forming an optical tweezer for single atoms. 
The tweezer spot is aligned to the ONF by monitoring the tweezer light scattered into the ONF guided modes. A blue-detuned laser light with a wavelength of 830 nm and power of around 1 mW, is launched into the tapered fiber to avoid the atoms sticking to the ONF surface. The background photons induced by the 830 nm laser and the tweezer light are filtered out using volume Bragg gratings, interference filters and filter cavities.  

The experiments are carried out by monitoring the photon counts through the ONF cavity while the tweezer beam is focused on the ONF and laser-cooled Cs-atoms are continuously loaded into the ONF-trap from a magneto optical trap (MOT). 
The typical photon counts measured through the ONF cavity, is shown in the left panel of Fig. 2(a). The right panel shows the histogram of the photon counts. It may be seen that discrete step-like signals with a height of 62$\pm$13 counts/ms are observed above a background of 17$\pm$7 counts/ms. The typical temporal duration of the step-like signal ranges from 10 to 100 ms. The signal resembles the single atom fluorescence signal measured in a conventional tightly focused dipole trap operating in collisional blockade regime \cite{grangier}. It disappears when the tweezer beam or the repump beam of the MOT is switched off. This clearly indicates that single atoms are trapped and interfaced to the ONF cavity. The efficient channeling of the trapped single atom fluorescence into the ONF guided modes enables the real-time observation of the step-like fluorescence signal. 

\begin{figure}[tbp]
\includegraphics[width=\linewidth]{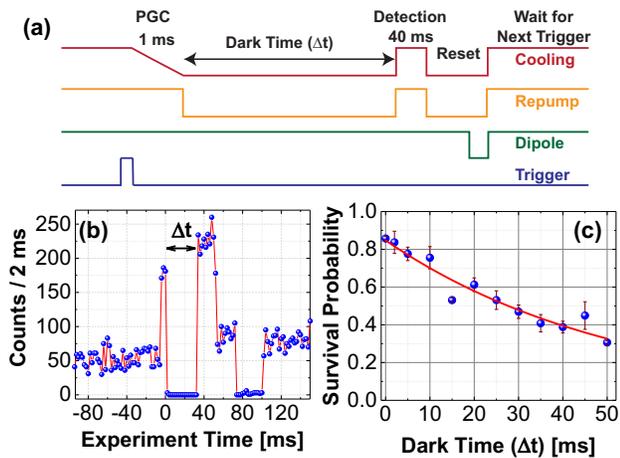}
\caption{\label{fig3} (Color Online) Lifetime of the ONF-trap. (a) Schematic of the experimental timing sequence. (b) A typical experimental event. The blue dots show the measured photon counts in each time-bin (2 ms) and the red trace is just a guide to eye. (c) The measured survival probabilities plotted as a function of different dark times ($\Delta t$). The blue dots show the measured data with error-bars and the red curve shows the exponential fit to the data.}
\end{figure}

We have carried out photon correlation measurements of the fluorescence signal to further clarify the atom number and the dynamics of the trap. The magenta traces in Fig. 2(b) show the typical normalized photon correlation signal ($g^2(\tau)$) measured through the ONF cavity. 
The correlation signal shows anti-bunching behavior with $g^2(0)\simeq 0.47$. The anti-bunching of the fluorescence signal, confirms that indeed single atoms are trapped on the ONF. It should be noted that $g^2(0)>0$ is due to the presence of background light scattered from the MOT beams into the ONF cavity. It may be seen that apart from the central anti-bunching dip the correlation signal shows a bunching behaviour in longer time-scales. In order to further understand the behaviour of the correlation signal, we fit the correlation signal using a model (see e.g. \cite{tapster, meschede}) given by
\begin{equation}
g^2(\tau)=1-(1+C_0)e^{-\frac{|\tau|}{t_0}}+C_1e^{-\frac{|\tau|}{t_1}}
\end{equation}
where $t_0$ and $t_1$ describes the characteristic time-scales of the anti-bunching and bunching signals, respectively. The coefficient $C_1$ describes the amplitude of the bunching signal and ($C_1-C_0$) gives the value of $g^2(0)$. The blue curve in Fig. 2(b) shows the fitting using the above model. From the fit, we estimate the parameters $t_0$, $t_1$, $C_0$ and $C_1$ to be 9.6$\pm$0.7 ns, 800$\pm$10 ns, 0.35$\pm$0.10 and 0.82$\pm$0.01, respectively. It should be noted that the $t_0$ may correspond to the cavity enhanced decay rate of the atom. On the other hand, the $t_1$ is due to the periodic modulation of the fluorescence signal. The motion of the atom in the trap may lead to change in the $\eta$ and hence the modulation of the signal.

The real-time observation of single atoms trapped on the ONF enables various experiments to be performed in a deterministic way. Once a single atom is trapped on the ONF, a step-increase in the photon counts is observed through the ONF. This step-like signal is used to trigger the experimental sequence. As a demonstration experiment, we present the lifetime measurement of the trapped single atoms on the ONF. The time sequence of the experiment is shown in Fig. 3(a). The experiment starts with polarization gradient cooling (PGC) for 1 ms, followed by a dark-period of $\Delta t$, during which the MOT beams are switched off. After the dark-period, the MOT beams are switched on again to detect the presence of the atom in the trap. 

A typical measurement event is shown in Fig. 3(b). It may be seen that the presence of a single atom in the trap was inferred from the sudden rise in the photon counts and the experiment was triggered at 0 ms time. There was a dark-period of 32 ms during which the photon counts were zero. A high level of photon counts after the dark-period indicates that the atom was still present in the trap. The blue dots in Fig. 3(c), show the survival probabilities for different dark times ($\Delta t$). The red curve shows the exponential fit. From the fit, the trap lifetime is estimated to be 52$\pm$5 ms.

\begin{figure}[tbp]
\includegraphics[width=\linewidth]{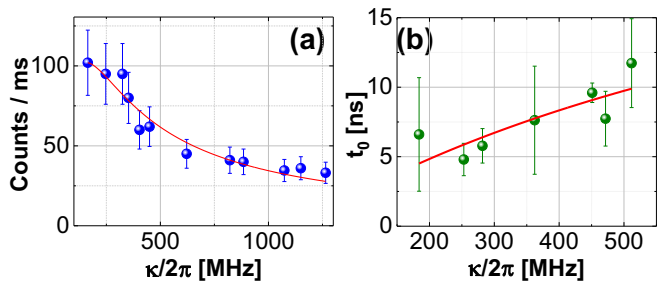}
\caption{\label{fig4} (Color Online) Photon statistics of the fluorescence signal for different cavity modes. (a) The measured fluorescence counts plotted against the linewidths ($\kappa/2\pi$) of the cavity modes. The blue dots show the measured data with error-bars and the red curve shows the theoretical fit to the data. (b) The estimated rise time of the antibunching signal ($t_0$) plotted against the cavity linewidths. The green dots show the measured data with error-bars and the red curve shows the theoretical fit to the data.}
\end{figure}

In order to estimate the cavity QED characteristics of the coupled atom-cavity system, we have analyzed the photon statistics of the fluorescence signal for different cavity modes. The measured fluorescence counts (height of the step-like fluorescence signal) are plotted against the linewidths ($\kappa/2\pi$) of the cavity modes in Fig. 4(a). It may be seen that the fluorescence counts increases with decreasing cavity linewidth. We fit the data using the following relation (see e.g. \cite{meschede, hinds})
\begin{equation}
n_p=\alpha\frac{\Omega^2}{\gamma_0 C}\frac{|C'|^2}{|1+C'|^2}
\end{equation}
where $C'=g_0^2/[(\kappa/2-i\Delta_c)(\gamma_0/2-i\Delta_a)]$ is the complex cooperativity parameter.
Here $g_0$, $\kappa$ and $\gamma_0$ ($=2\pi \times 5.2$ MHz), are the atom-cavity coupling rate, the cavity decay rate and the decay rate of atom in free space, respectively. $\Omega$, $\Delta_c$ ($=0$ MHz) and $\Delta_a$ ($\simeq -3\gamma_0$) are the driving field's (MOT cooling beams) Rabi frequency and detunings from the cavity and atomic resonances, respectively. $C=4g_0^2/(\kappa \gamma_0)$ is the on-resonance single atom cooperativity. The experimental detection efficiency is described by the proportional factor $\alpha$. From the fit, the atom-cavity coupling rate is estimated to be $g_0/2\pi=34\pm2$ MHz. It should be noted that $\alpha$ and $\Omega$ are just proportional factors and assuming the experimentally measured detection efficiency $\alpha=4$\%, the driving Rabi frequency is estimated to be $\Omega/2 \pi=5.4\pm0.2$ MHz. From the $g_0$ value, we estimate a cooperativity of $C=5.4\pm$0.6 and a cavity enhanced channeling efficiency ($\eta_c=(P_0\eta+C)/(P_0+C)$) of 85$\pm$2\% for the cavity mode having a linewidth of 164 MHz, where $P_0\simeq 1.044$ is the Purcell factor due to the nanofiber in the absence of the cavity \cite{famsan1,CPIreview, famsan2}. 

We have also analyzed the photon correlation signal for different cavity modes. Figure 4(b) shows the estimated rise time of the antibunching signal ($t_0$) plotted against the cavity linewidths. It may be seen that $t_0$ increases with increasing cavity linewidth. We fit the data using a relation
$t_0=1/(P_0\gamma_0+4g_0^2/\kappa)$ \cite{CPIreview, famsan2}. From the fit, the atom-cavity coupling rate is estimated to be $g_0/2\pi=36\pm$1 MHz, which is in reasonable agreement with analysis of Fig. 4(a).  

It should be noted that the selection of ONF diameter is an essential requirement for the trapping scheme. The typical ONF diameter of 400-500 nm widely adapted in reported experiments may not be suitable for the present trapping scheme. Due to proper selection of ONF diameter, the single atom fluorescence can be observed in real-time above the scattering background from the MOT beams. A similar trapping scheme have been reported for silicon-nitride based nanobeam cavity \cite{lukin}. However, in that scheme single atoms are observed in the free-space tweezer and then transported to the nanobeam cavity. Such moving tweezer based trapping scheme may not be suitable for ONF cavities having much narrower linewidths as the tweezer beam itself can induce shift in the cavity resonance due to strong photo-thermal effects \cite{kpn2}. Moreover, the scheme may be seriously affected by the position stability of such a long tapered fiber. 

The data presented in Figs. 2 and 3, were measured for a cavity mode having a linewidth of 450 MHz. The observed fluorescence counts and the $t_0$ are well understood from the analysis of Figs. 4(a) and (b). However, it must be noted that from the estimated cavity length, the atom-cavity coupling rate is estimated to be $g_0=(cP_0\gamma_0 \eta/L)^{1/2}= 2\pi \times 63 \pm 3$ MHz, where $c$ is the speed of light in vacuum \cite{CPIreview, famsan2}. This may be understood from the fact that the confinement of the trap along the axis of ONF is weak and due to the motion of the atom in the trap it can transit over few nodes and antinodes of the cavity mode. As a result, the $g_0$ estimated from the photon statistics, is an averaged value. This is also evident from the bunching behaviour of $g^2(\tau)$ resulting from the periodic modulation of the fluorescence signal. The axial confinement can be further improved by introducing a blue-detuned standing wave into the ONF guided mode.        
The trap lifetime is similar to previously reported values for ONF based guided mode traps \cite{arno1,aoki}. The lifetime is mainly limited by the mechanical modes of the tapered fiber \cite{arno3} and can be improved using Raman-cooling techniques \cite{arno4}.

In summary, we have demonstrated that fluorescence of individual single atoms trapped on the ONF cavity, can be readily observed in real-time through the fiber guided modes. The trap lifetime was measured to be 52$\pm$5 ms. From the photon statistics, the effective coupling rate of the atom-cavity interface was estimated to be 34$\pm$2 MHz. This yields a cooperativity of 5.4$\pm$0.6 and a cavity enhanced channeling efficiency of 85$\pm$2\% for a cavity mode having a linewidth of 164 MHz. These results may open new possibilities for building trapped single atom based quantum interfaces on an all-fiber platform.

\begin{acknowledgments}
We acknowledge Kohzo Hakuta, Tetsuo Kishimoto and Makoto Morinaga for fruitful discussions. This work was supported by the Japan Science and Technology Agency (JST) as one of the Strategic Innovation projects. KPN acknowledges support from a grant-in-aid for scientific research (Grant no. 15H05462) from the Japan Society for the Promotion of Science (JSPS) and support from Matsuo Science Promotion Foundation, Japan.

K. P. Nayak and J. Wang contributed equally to this work.

$^\star$ kalipnayak@uec.ac.jp     $^\ddagger$ wangjie605@126.com

\end{acknowledgments}





\end{document}